\documentclass[12pt,letterpaper,onecolumn]{article}
\usepackage[latin1]{inputenc}
\usepackage{amsmath}
\usepackage{amsfonts}
\usepackage{amssymb}
\usepackage{braket}
\usepackage{enumerate}

\usepackage[left=1.00in, right=1.00in, top=1.00in, bottom=1.00in]{geometry}

\author{\\
\\
Horace P. Yuen\\
Department of Electrical Engineering and Computer Science\\
Department of Physics and Astronomy\\
Northwestern University, Evanston Il. 60208\\
yuen@eecs.northwestern.edu
}
\title{PROBLEMS OF EXISTING UNCONDITIONAL SECURITY PROOFS IN QUANTUM KEY DISTRIBUTION}

\begin{document}
\linespread{1}
\maketitle
\linespread{2}
\begin{abstract}
It is repeatedly and persistently claimed in the literature that a
specific trace criterion $d$ would guarantee universal composition security
in quantum cryptography. Currently that is the sole basis of unconditional
security claim in quantum key distribution. In this paper, it is shown
that just security against known-plaintext attacks when the generated
key is used in direct encryption is not provided by $d$.
  The problem is directly connected with several general problems in the existing unconditional security
proofs in quantum key distribution. A number of issues
will be clarified concerning the nature of true security, privacy amplification,
key generation rate and the mathematical approach needed for their
determination in concrete protocols.

Note added: It has now been shown that security for a proper form of $d$ is obtained in an average sense. See arXiv:1205.5065v2. Many issues in this paper have been clarified and further elaborated, see arXiv:1201.2804v1 and references cited therein.
\end{abstract}
\maketitle

\section{Introduction and Summary}

In quantum key distribution (QKD) there have been many security proofs
offered on the ``unconditional security''
of various protocols of the BB84 variety. (For a recent review
see ref {[}1{]}.) Until 2004-2005 and in many papers till the present
day, the security criterion adopted is the attacker Eve's quantum
accessible information ($I_{ac}$) on the generated key $K$, which
is the maximum mutual information Eve has on $K$ from a measurement
result on her probe she may set during the key generation process.
Security of $K$ before it is actually used is called ``raw
security'' {[}2{]}, to distinguish it with composition
security when $K$ is actually used in an application for which Eve
may possess additional information related to $K$. In particular,
when $K$ is used for encryption, part of $K$ may be known to Eve
in a known-plaintext attack (KPA) to help her get at the rest of $K$.
KPA takes a particularly simple form when $K$ is used in the often
suggested one-time pad format.

While ``universal'' composition security
is a complicated matter which is perhaps not needed in its full generality,
KPA security is necessary because that is one main weakness of conventional
symmetric key ciphers QKD purports to overcome. Indeed, there is otherwise
no need for QKD since its raw security is worse than that of conventional
ciphers in which the key is also typically totally hidden by uniformly
random data {[}2{]}. In this paper all security under discussion is
information-theoretic (IT), and symmetric key cipher is the proper
comparison with QKD, not purely complexity-based cipher such as RSA.
This is because a shared message authentication key is necessary in
QKD during key generation, and in any case a short shared secret key
can always be employed.

It was claimed in {[}3{]} that an exponentially small $I_{ac}$ (for
an $n$-bit key $K$) would guarantee universal composition security
and that is applicable to most previous security proofs. The claim
was established through an inequality between $I_{ac}$ and a trace
distance criterion that has been given by the notation ``$d$''
in many papers since {[}4{]}, an abbreviation we adopt in this paper.
This $d$ is supposed to give the trace distance between the states of an ``ideal''
protocol and the ``real'' protocol under
Eve's attack. It was shown in {[}4{]} through an explicit construction
involving quantum information locking that exponentially small $I_{ac}$
does not imply KPA security, specifically the last bit of $K$ may
be leaked deterministically when $n-1$ bits of $K$ are known to
Eve from a KPA. This small leak has been enlarged to a ``spectacular
failure'' of the $I_{ac}$ guarantee, under which it
is possible that ``leakage of even a logarithmic number
of key bits compromises the secrecy of all the others''.
{[}5{]}

The remedy, according to {[}4-7{]}, is to use the criterion $d$ directly.
Indeed, $d$ is the \textit{only} basis of QKD unconditional security
claim at present including any use of privacy amplification {[}1,
footnote 20{]}, {[}6-7{]}. There are three different ``interpretations''
on what $d\leq\epsilon$ asserts, each of which is claimed to imply
universal composition security. We would concentrate on KPA security
in this paper, which is much simpler and can be treated directly.
The three interpretations are:

\begin{enumerate}[(i)]
\item  ``The real and the ideal setting can be considered
to be identical with probability at least $1-\epsilon$''.
{[}6{]}
\item The parameter $\epsilon$ can be understood as the ``maximum
failure probability'' of the real protocol, i.e., the
maximum probability that the real protocol ``deviates
from the behavior of the ideal protocol''. {[}8{]}
\item ``Distinguishability advantage'' between
the real and the ideal protocols is bounded by $\epsilon$. {[}3{]}
\end{enumerate}

In this paper all three interpretations will be analyzed, only briefly
on (i) because {[}2{]} already shows that (i) cannot be true.
With (ii) interpreted with respect to a specific scenario
so that it is different from the more general (i), it is refuted by
a specific KPA counter-example. In fact, $d$ could be interpreted
as the difference between two probabilities but it does \textit{not}
have a probability interpretation itself. We will explain why (iii)
does not lead to KPA security in general.
A different criterion $d^{\prime}$ is needed for such interpretation.

In sum, there is no QKD unconditional security proof at all against
attacks with quantum memory.  The ramification of this security
failure will be elaborated. The actual QKD security situation will
be discussed in regard to the secure key generation rate, privacy
amplification, and the necessity of using $M$-ary quantum detection
theory in quantifying fundamental security performance unless $d^{\prime}$ is bound.

\section{CLASSICAL VARIATIONAL DISTANCE AND QUANTUM TRACE DISTANCE}

The classical variational distance $v(P,Q)$ between two probability
distributions $P=\{p_{i}\}$ and $Q=\{q_{i}\}$ on the same sample
space is given by {[}9{]}

\begin{equation}
v(P,Q)=\frac{1}{2}\sum_{i}\mid p_{i}-q_{i}\mid\end{equation}

\noindent with $0\leq v\leq1$. The quarantee $v\leq\varepsilon$
is equivalent to, for any event $\mathcal{E}$, the probabilities
of $\mathcal{E}$ from $P$ and $Q$ satisfy

\begin{equation}
\mid p(\mathcal{E})-q(\mathcal{E})\mid\leq2\varepsilon\end{equation}

\noindent Indeed we have {[}9, p.299{]}

\begin{equation}
2v(P,Q)=\max_{\varepsilon}\mid p(\mathcal{E})-q(\mathcal{E})\mid\end{equation}

An important case for our purpose is when $Q$ equals the uniform
distribution $U$, $u_{i}=1/N$ for sample space of size $N=2^{|K|}$
while Eve has distribution $P$ for $K$. Then (2) shows

\begin{equation}
\mid p(\tilde{K})-\frac{1}{2^{|\tilde{K}|}}\mid\leq 2\epsilon\end{equation}

\noindent for any subset $\tilde{K}$ of $K$ when averaged over the possible value of $\tilde{K}$. (Note added: such average (4) is established in arXiv:1205.5065v2, the original statement (4) is not correct for individual $\tilde{K}$ value.) However, when $\frac{\varepsilon}{2^{-n}}\gg1$,
$P$ may be very different from $U$ in regards to the possible $p(\mathcal{E})$
even when $\varepsilon$ is exponentially small, say $\varepsilon=2^{-n/2}$.
Whether something is small in a cryptographic context has to be judged
with respect to the key length or data length with exponentiation
if appropriate.

The quantum trace distance between two density operators $\rho$ and
$\sigma$ on the same state space is

\begin{equation}
D(\rho,\sigma)=\frac{1}{2}\parallel\rho-\sigma\parallel_{1}\end{equation}

\noindent with $0\leq D\leq1$. It can be readily shown that $D(\rho,\sigma)\leq\varepsilon$
implies $v(P,Q)\leq\varepsilon$ for any quantum measurement which
gives $P$ and $Q$ from $\rho$ and $\sigma$ {[}10{]}. By using
the basis that diagonalizes $\rho-\sigma$, $D(\rho,\sigma)$ itself
can be achieved by a measurement in the form $v(P,Q)$. Thus, we have
the equivalence of variational distance with trace distance as a criterion.

It is important to stress that $D(\rho,\sigma)\leq\varepsilon$ does
\textit{not} imply that $\rho$ and $\sigma$ are close, similar to
$v(P,Q)\leq\varepsilon$ does \textit{not} imply $P$ and $Q$ are
close, unless $\varepsilon$ is small enough. Incorrect understanding
of the security situation would result if the quantitative level of
$\varepsilon$ relative to $2^{-n}$ is not attended to for an $n$-bit
or $n$-qubit sequence. This is due to the large freedom of $P$,
in particular $p_{1}$, that is possible under such a constraint for
fixed $\sigma$ or $Q$. This has been emphasized in {[}2,11-13{]}.

\section{PROBLEM FORMULATION AND RAW SECURITY GUARANTEE }

During the key generation process, Eve sets her probe and the protocol
goes ahead after intrusion level estimation.
We assume that everything goes well on the user's end.
At whatever time when
Eve measures on her probe, she would obtain a \textit{whole} probability
distribution on correctly estimating the different possible values
of $K$ {[}13{]}. Classically the quantitative raw security problem
can be formulated as follows. We will use upper case letter for a
random variable (vector) with its specific value denoted by the corresponding
lower case letter.

Let $X$ be an m-bit data sequence random variable picked by user
\textit{A} and $Y$ Eve's observation random variable of any possible
length and alphabet size. The transition probability $p(y|x)$ and
a priori distribution $p(x)$ are fixed by the cryptosystem and chosen
attack. The user $B$ observes the random variable $Z$ specified by
the cryptosystem, applies an openly known known error-correcting code
(ECC) to get a data estimate $\hat{X}(Z)$ which is presumably error
free, and then an openly known privacy amplification code (PAC) to
yield a final generated key $K$. The ECC and PAC can be combined
to yield directly $K(Z)$. From $Y$ Eve forms her estimate $K(Y)$.
The timing of Eve's knowledge of various openly known codes is implicit
in the possible $p(y|x)$ she could obtain.

With Bayes rule and the known ECC+PAC, Eve forms from $y$ the conditional
probability distribution (CPD) on $K$, $p(k|y)$, which gives Eve's
success probability of getting the entire $k$ for each possible value
of $K$. We will use $P=\{p_{i}\}$, $i\in\{1,\ldots,N\}$, for this
CPD, suppressing the dependence on $y$. Any single-number criterion
on $K$, be it mutual information or variational distance, merely
expresses a constraint on $P$. The Markov Inequality [9] can be used
to convert an average constraint to an individual one for a nonnegative
random variable, here it is $p(k|Y)$ for each $k$ and random $Y$.
We order $p_{i}$ so that $p_{1}\geq p_{2}\geq\ldots\geq p_{N}$.
Thus, $p_{1}$ is Eve's optimal probability of estimating $K$ correctly
given $y$. It is a most significant number concerning the security
of $K$, as we will see.

With $I(K;Y)$ denoting the mutual information between any two random
variables $K$ and $Y$, we use the following notations

\begin{equation}
\delta_{E}\equiv v(P,U),\;\;\; I_{E}\equiv I(P;U)\end{equation}

\noindent For simplicity, we take the data $X$ to be uniformly distributed
and the same for $K$ obtained fom it via ECC+PAC, the ideal situation.
Thus $\delta_{E}$ and $I_{E}$ in (6) are indeed single-number measures
of Eve's ``information'' on $K$.

Note that it is not sufficient to employ a criterion that would give
perfect IT security when it has its limiting value, say $\delta_{E}=0$
or $I_{E}=0$, but using it for a relatively large nonzero value.
The issue is a quantitative one and whether the security guarantee
is adequate depends on the exact value that can be obtained in a concrete
protocol, as we will see.

We have shown in {[}11-13{]} that for $I_{E}=2^{-l'}, l'>0,$ Eve's maximum
probability of getting the whole $K$ can be as big as

\begin{equation}
p_{1}\sim2^{-l},\;\;\;\;\;\;\; l=l^{'}+\log n\end{equation}

\noindent Unless $l\sim n$, the raw security guarantee of $I_{E}\leq2^{-l'}$
is very far from that of a uniform key. The subsets of $K$ suffer
similarly {[}13{]}. When $l^{'}$ approaches $n-\log n$, more exact
estimate of $p_{1}$ {[}16{]} needs to be used in lieu of $2^{-l}$
since $l$ cannot exceed $n$. The practical experimental value of $l'\sim21$
for $n\sim4000$ {[}14, 15{]} is quite an inadequate guarantee, especially
after the application of Markov Inequality {[}2,13{]}. Generally,
{}``exponentially small in $n''$ can be very misleading because
the rate $\lambda$ in $l=\lambda n$ is the real crux of the security
situation. We will see this repeatedly in the following.

The $\delta_{E}$ guarantee suffers a similar problem {[}2,13{]} because
for $\delta_{E} \le 2^{-l}$, the averaged (over Y) $\overline{p}_{1}$
can be as big as

\begin{equation}
\overline{p}_{1}=2^{-l}-\frac{1}{N}\:\:\:\: with\:\:\:\:\delta_{E}=2^{-l}\end{equation}

\noindent Thus, unless $l\sim n$ as indicated after (4) above, a
$\delta_{E}\leq\epsilon$ raw security guaranree is not really better
than that of $I_{E}\leq\epsilon$.

\section{INCORRECTNESS OF INTERPRETATIONS $(i)$ AND $(ii)$}

Let $\rho_{E}^{k}$ be Eve's probe state when $K$ has value $k$
with probability $p_{0}(k)$ before B measures, $p_{0}(k)=U$ in the
ideal case. The possible $\rho_{E}^{k}$ are limited by the users' intrusion level estimation. Let

\begin{equation}
\rho_{K}\equiv\sum_{k}p_{0}(k)\mid k\rangle\langle k\mid\end{equation}

\noindent be the $p_{0}(k)$-mixed state on $N$ orthonormal $\mid k\rangle's$.
Let $\rho_{E}$ be the $K$-averaged state, and $\rho_{KE}$ the joint
state

\begin{equation}
\rho_{E}\equiv\sum_{k}p_{0}(k)\rho_{E}^{k}\end{equation}

\begin{equation}
\rho_{KE}\equiv\sum_{k}p_{0}(k)\mid k\rangle\langle k\mid\otimes\rho_{E}^{k}\end{equation}

\noindent The criterion $d$ is defined to be,

\begin{equation}
d\equiv\frac{1}{2}\parallel\rho_{KE}-\rho_{K}\otimes\rho_{E}\parallel_{1}\end{equation}

\noindent A key satisfying $d\leq\varepsilon$ is called ``$\epsilon$-secure''
by defintion {[}6{]} in the case $p_{0}(k)=U$.

The ``lemma 1'' of {[}6{]} and {[}16{]}
was given the following interpretation on $v(P,Q)\leq\varepsilon$:
the two random variables $K$ and $Y$ described by $P$ and $Q$
take on the same value with probability $\geq1-\varepsilon$. (The
lemma says there exists a joint distribution of $K$ and $Y$ which
gives this result. That joint distribution is in fact the optimal
one for this interpretation.) Given this incorrect interpretation
as premise, it can be validly deduced {[}6, p.414{]},{[}2{]} that under $d\leq\varepsilon$,
``the real and the ideal setting can be considered to
be identical with probability at least $1-\varepsilon$''.
As discussed in {[}2,13,17{]}, this interpretation of $d$ is not
a consequence of ``lemma 1'' in {[}6{]}
or {[}16{]} but an incorrect interpretation of that lemma 1. We may
note here that there is no physically meaningful joint distribution
that gives $P$ and $Q$ as marginals other than the product distribution
$PQ$ which applies in this situation. Thus, the two random variables
$K$ and $Y$ would take the same value only as a result of random
collision with probability $1/N$, $N$ the size of the sample space,
even when $P$ and $Q$ are the same distribution. As concluded in
{[}2{]}, interpretation (i) is simply false and not just unproven.

Going onto interpretation (ii), observe that its wording in {[}8{]}
is very ambiguous. It can mean either interpretation (i), or (iii)
with $\varepsilon$ as the probability \textit{difference} between
the real and the ideal cases. We would give this ``failure
probability'' a distinct literal interpretation from
the words of {[}8{]}, since it is the sole basis of the QKD unconditional
security claim in the recent review {[}1{]}. In lieu of random variable
identity or coincidence of (i), we restrict (ii) to apply just to
specific KPA scenarios in which performance can be readily quantified.
The following simple counter-example shows such interpretation (ii)
cannot be expected to hold, not just unproven.

Consider the following simplest information locking example, for a
two-bit $K$ with $\rho_{E}^{k}=\rho^{k_{1}}\otimes\rho^{k_{2}}$
for the two bits $k_{1}$ and $k_{2}$. Let $|i\rangle$, $i\in\overline{1-4}$,
be the four BB84 states on a qubit, with $\langle1|3\rangle=\langle2|4\rangle=0$.
Let $\textrm{P}_{i}$ be the projectors into $|i\rangle$, and

\begin{equation}
\begin{aligned}
\rho_{E}^{11}=\frac{1}{2}(\textrm{P}_{1}\otimes \textrm{P}_{1}+\textrm{P}_{3}\otimes \textrm{P}_{2})\\
\rho_{E}^{10}=\frac{1}{2}(\textrm{P}_{1}\otimes \textrm{P}_{3}+\textrm{P}_{3}\otimes \textrm{P}_{4})\\
\rho_{E}^{01}=\frac{1}{2}(\textrm{P}_{2}\otimes \textrm{P}_{1}+\textrm{P}_{4}\otimes \textrm{P}_{2})\\
\rho_{E}^{00}=\frac{1}{2}(\textrm{P}_{4}\otimes \textrm{P}_{3}+\textrm{P}_{4}\otimes \textrm{P}_{4})\\
\end{aligned}
\end{equation}

\noindent Thus, $k_{2}$ is locked into the second qubit through $k_{1}$,
and is unlocked by measuring on the 1-3 or 2-4 basis given the knowledge
of $k_{1}$. This $\rho_{E}^{k}$ does not yield a $\rho_{E}= I/4 $, but since $d$ is equal to {[}6, lemma 2{]}
$ $\begin{equation}
d=\frac{1}{2}E_{K}[\parallel\rho_{E}^{k}-\rho_{E}\parallel_{1}],
\end{equation}
let us evaluate ``ideal'' comparison $\parallel\rho_{E}^{k}-I/4\parallel_{1}$ which is easily computed to be $1/2$.
However,
knowing $k_{1}$ implies $k_{2}$ is compromised for sure, not with
a maximum failure probability $1/2$, contradicting the interpretation
(ii) in this specific situation. (Note that the $d\leq\varepsilon$
guarantee is supposed to apply to any $\rho_{E}^{k}$ in (12)).

Indeed, any locking information scenario provides a counter-example to (ii) similar to the example of (13).
 Let $I_{ac}= 2^{-l'}$ and with a $\rho_{E}^{k}$ that leaks the rest of $K$ from its $l$ bits according to (7). The corresponding $d$ must be less than $1$ since $D(\rho,\sigma)=1$ if and only if $\rho$ and $\sigma$ have orthogonal ranges.

It does \textit{not} appear there is any probability meaning one can
sensibly give to $\varepsilon$ in $d\leq\varepsilon$, which is just
a numerical measure of the difference between two density operators
similar to $\varepsilon$ between $P$ and $Q$ in $v(P,Q)\leq\varepsilon$.
There is simply no basis to assign probability distribution to the
security situation after all the parameters are fixed, i.e., there
is no more random system parameter that could give rise to such probability
distribution. The incorrect probability interpretation of $\varepsilon$
in $v(P,Q)\leq\varepsilon$ is responsible for the incorrectness of
interpretation (i). Here we see that any probability interpretation
of $\varepsilon$ itself, whatever the ``failure probability''
may be, would fail similarly.

\section{SECURITY FAILURE UNDER INTERPRETATION $(iii)$}

Going on to interpretation (iii), note the huge difference between it and the other two interpretations. According to (i) and (ii), $d=1/2$ in the example of (13) is the probability Eve can succeed, which is not true. According to interpretation (iii), Eve may succeed at $1/2+d=1$, which is true (we actually use $d'$ to get the ideal situation.)

It may be observed that $\varepsilon$
in $D(\rho,\sigma)\le\varepsilon$ is \textit{not} the success probability
of distinguishing $\rho$ from $\sigma$ by a measurement. That is
given by the well known {[}18{]} probability ${P_{c}}$ of correct decision,

\begin{equation}
P_{c}=\frac{1}{2}+\frac{1}{2}D(\rho,\sigma)\end{equation}

\noindent Note that Eve is not usually trying to make a binary decision
with her attack. A major source of confusion may have arisen from
calling $\rho$ and $\sigma$ ``$\epsilon$-indistinguishable''
when $D(\rho,\sigma) \leq \varepsilon$, as if $\rho$ and $\sigma$ can
only be distinguished with probability $\sim\varepsilon$. Actually
just (15), or (2) for any measurement, is the mathematical statement
of ``$\epsilon$-indistinguishable''. Any
other claimed consequence needs to be mathematically \textit{expressed
and derived} from this mathematical given. Such development has \textit{not}
been provided for universal composition security (which is different
from ``composability'' of the criterion),
not just for KPA. Before going into the KPA issue we will show
that (iii) is not generally true even
for raw security, because the ``ideal''
situation is not captured by $\rho_{U}\otimes\rho_{E}$, as follows.

Consider a term $\parallel\rho_{E}^{k}-\rho_{E}\parallel_{1}$ in (14) apart from averaging over K. It implies

\begin{equation}
v(p(y|k),p(y))\leq2\varepsilon\end{equation}

\noindent for Eve's observation $Y$ on her probe. Notice that under (iii), $\epsilon$ is merely a single-number quantitative
measure of difference, and thus has much weaker meaning than the equality
of whole state or distribution. Indeed, it is clear that the level
of $\epsilon$ becomes crucial, as we saw in section II, even if it is measured
with respect to the ``ideal''.

The following different criteria $d'$ should be used for security proofs (arXiv: 1205.5065v2)
\begin{equation}
d \equiv \frac{1}{2} \sum_{k} ||p_o(k)\rho_E^k - \frac{1}{N}\rho_E||_1
\end{equation} Note that there is no freedom of choice in replacing $\rho_E$ of (17) by another state.

The distributions that lead to (7) appear to be of the form suitable for information locking with small $I_{E}$. Indeed, in [5] there is an $l^{'}$ factor in addition to $O(\log n)$ in the expression for the number of
unlocking bits in the key segment, exactly as in (7). In any case,
complete raw and composition security would obtain if the required number of bits to increase $p_{1}$
to 1 goes up to $n$. For $I_{ac}\le2^{-l^{'}}$, this would happen at

\begin{equation}
l^{'} \ge n- \log n \end{equation}

\noindent For the case $d'\le2^{-l}$ or $\delta_{E}\le2^{-l}$, it would happen at

\begin{equation}
l\ge n\end{equation}

It may be observed that KPA may significantly lower $p(\tilde{K})$, Eve's success probability of getting any $\tilde{K}\subseteq K$, to an
unacceptable level without deterministically compromising the whole $K$. Partial information locking of $K$ must, therefore, be dealt with also in a
fundamental security analysis.
The KPA case alone already shows that there is no universal composition security guarantee from $d$, at least when it is below a certain quantitative level.
This is in fact a problem of any single-number criterion, but we will not go into the general issue in this paper.

\section{RELATION BETWEEN HOLEVO QUANTITY $\chi$ AND THE CRITERION $d$}

The classical form of $d$, say as obtained from a measurement, is

\begin{equation}
\delta = \frac{1}{2}v(p(y|k)p_{0}(k);p(y)p_{0}(k))\end{equation}

\noindent The following simple relation between the above $\delta$ and the classical
mutual information $I(K;Y)$ is an immediate consequence of the well
known {[}9, p.300{]} relation between relative entropy and variational
distance by considering $p(y,k)$ relative to $p(y)p_{0}(k)$. We
have

\medskip{}

\medskip{}

\noindent Lemma 1:

\noindent The $\delta$ of (16) is upper bounded by $I(K;Y)$ in the
form

\noindent \begin{equation}
2\delta^{2}\leq I(K;Y)\end{equation}

\medskip{}
\medskip{}

\noindent From (21), one obtains the weak bound,

\medskip{}

\medskip{}

\noindent Lemma 2:

\noindent The criterion $d$ is upper bounded by the quantum accessible
information $I_{ac}$ that Eve can get from her probe

\noindent \begin{equation}
2d^{2}\leq2^{|K|}I_{ac}\end{equation}

\medskip{}
\medskip{}

\noindent Proof:

From (14) each term, $\parallel\rho_{E}^{k}-\rho_{E}\parallel_{1}$ ,
is bounded by a measurement result $Y^{(k)}$ satisfying (21). Thus,

\begin{equation}
d\leq E_{K}[\frac{I(K=k;Y^{(k)})}{2}]^{\frac{1}{2}}\end{equation}

\noindent By Jensen's Inequality,

\begin{equation}
d\leq[E_{K}\frac{I(K=k;Y^{(k)})}{2}]^{\frac{1}{2}}\end{equation}

\noindent which is bounded as (22) by adding many nonnegative terms for each $Y^{(k)}$ inside the
$[.]^{1/2}$ of (24) to get $\sum_{k}I(K;Y)=2^{|K|}I(K;Y)$ .

\medskip{}
\medskip{}

It is on the basis of  equ(16) in {[}3{]}, which is equivalent to (22),
that the incorrect conclusion is drawn in {[}3{]} that exponentially
small $I_{ac}$ would guarantee composition security in previous security
proofs. Our (18) or (22) shows that the exponent needs to be nearly all
of $n$ for such conclusion to hold. In the case of {[}14{]} with
$n\sim4000$, this means the exponent needs to be as big as $l^{'}\sim3880$.

The proper quantum generalization of Lemma 1 is not Lemma 2 but the
following

\noindent Lemma 3:

\noindent The Holevo quantity {[}10{]},

\begin{equation}
\chi=S(\rho_{E})-E_{K}[S(\rho_{E}^{k})]\end{equation}

\noindent bounds $d$ in the form,

\begin{equation}
2d^{2}\le\chi\end{equation}

\noindent Proof:

Similar to the classical (21), (26) follows from theorem 5.5 of {[}19{]}
with the quantum relative entropy $S(\rho\parallel\sigma)$ for $\rho=\rho_{E}^{k}$
and $\sigma=\rho_{K}\otimes\rho_{E}$.

\medskip{}

Since the security criterion is supposed to work for each
and every $\rho_{E}^{k}$, consider the one that leads to $\chi$ insecurity. Let $\chi=2^{-2m}$, $m>0$. Thus from (26), $d\le 2^{-m}$ but it is insecure. However, since $I_{ac} \leq \chi$, an $I_{ac}$ insecurity does not imply a $\chi$ insecurity. Although $I_{ac}/n= \chi/n$ asymptotically [20,21], the total $I_{ac}$ and $\chi$ are not
necessarily close for large $n$. One scenario that is the case is when blocks of such $n$ bits are repeated $n'$ times themselves for large $n'$, which is however not realistically applicable to concrete protocols.

The next theorem shows that $\chi$ and $d$ have similar exponential behavior, and thus are actually similar security criteria.

\noindent Theorem 1:\\
\noindent Let $h(\cdot)$ be the binary entropy function.
Then
\begin{equation}
2d^{2}\leq \chi\leq8dn+2h(2d)\end{equation}

\noindent Proof: \\
The lower bound is Lemma 3. The upper bound is an immediate consequence of the
theorem in {[}22{]},  again using $S(\rho\|\sigma)$ for  $\rho=\rho_{E}^{k}$ and $\sigma=\rho_{K}\otimes\rho_{E}$.

With $d=2^{-l}$ and $\chi=2^{-l^{''}}$, the exponents when non-negative are related from (27) for $n \ge l$ as follows,
\begin{equation}
l-\log n - 4 \leq l^{''} \leq 2l\end{equation}
Basically (27)-(28) shows that the exponents of $d$ and $\chi$ for almost any $n$ are within a factor of two.

\section{IT SEMANTIC SECURITY AND PRIVACY AMPLIFICATION}

Note added: The term ``semantic security,'' which is also used in [2], is inappropriate since it has been used in a different sense in classical cryptography.

What kind of raw security guarantee on $K$ one should have that is comparable
to that of a uniform key? One can introduce the notion of information
theoretic semantic security directly as

\begin{equation}
|p(\tilde{K})-\frac{1}{2^{|\tilde{K|}}}|\leq\epsilon(\tilde{K})\end{equation}

\noindent where $\tilde{K}$ is any subset of $K$ and $\epsilon(\tilde{K})$ is
allowed to vary depending on $\tilde{K}$ in contrast to (4). Such ``semantic
security'' in complexity-based cryptography has been
developed extensively {[}23{]} and generalized in an IT context {[}24{]}. However, in the context of IT physical cryptography
in noise we should dispense with any algorithm in the definition and
consider the full correlated statistical behavior of the system model.
Thus, (29) expresses the direct comparison with the ideal uniform key.
In fact, as long as $\epsilon (\tilde{K})$ is small enough, such as $\epsilon (\tilde{K}) = 2^{-|\tilde{K}|}$, we would not need to require
that it can be driven to zero. It would be quite adequate, e.g., if
$\epsilon$ is a constant $=2^{-n}$ for an $n$-bit $K$ as discussed in section II.

In the case one can guarantee only $\epsilon=2^{-m}$ for
$m<n$, it follows immediately that no IT semantically secure key
(with arbitrarily small $\epsilon$) can be obtained by any further
processing on $K$ which is longer than $m$. This is an immediate consequence
of the fact that $p_{1}$ cannot be improved by any known deterministic
transformation on $K$. Indeed, the original $p_{1}$ that results from
Eve's measurement $Y$ before ECC+PAC also cannot be improved with such codes,
i.e., not by the transformation from $Y$ to her estimate of $K(Y)$. Thus,
the IT semantically secure key rate is reduced from the nominal one by a factor $m/n$.

Let us consider the security of such an $m$-bit key $K_{r}$ derived from the $n$-bit $K$. When it is obtained from a
$d'$ guarantee of (17), all the subset probabilities $p(\tilde{K})$ Eve may get by any measurement is properly bounded from (2).
The users can
guarantee $\frac{I_{ac}}{n} \le 2^{-m}$ for sufficiently large $m$ under any $\rho_{E}^{k}$ Eve can launch that passes intrusion level estimation. However, in this case the resulting
$p(\tilde{K})$ or $p(\mathcal{E})$ bound for $\tilde{K}$ would not be quantum mechanically fundamental because Eve could attack a specific subset $\tilde{K}$ of $K$
by an optimal measurement directed toward that subset instead of the whole $K$. Thus she has a $2^{|\tilde{K}|}$-ary detection problem instead
of a $2^{|K|}$-ary one. Specifically, consider $K$ in two parts $K_{1}, K_{2}$  $(K=K_{1}K_{2})$. In a KPA knowing $K_{1}=k_{1}$, the state to Eve is $\rho_{E}^{k_{1}K_{2}}$ and she has a $2^{|K_{2}|}$ -ary quantum detection problem instead of the original $2^{|K|}$ -ary one.  Her optimum $2^{|K_{2}|}$ -ary quantum detection performance cannot in general be obtained from the  $2^{|K|}$ -ary performance and subsequent classical reduction to  $2^{|K_{2}|}$ -ary case. Quantum mechanically there is no complete
measurement which covers all such possibilities while maintaining
performance, but there is classically.

\par The essential point is that quantum detection theory {[18, 25]} is the proper approach here for optimal
performance analysis, not ``information theory''
in the narrow sense. Thus, the raw security guarantee on $p(\tilde{K})$ we
have discussed is also not ultimate either except for the total $K$ itself.
Note that the effect of PAC on fundamental quantitative security also
needs to be ascertained by quantum detection theory. The alternative is to bound $d'$.

\section{Problems of Unconditional Security Proofs in QKD}

Note added: For updated criticisms see arXiv: 1205.3820v2 and arXiv: 1201.2804v1.

The criterion $d$ fails to provide KPA security and composition security in general.
If $I_{ac} \le \varepsilon = 2^{-l}$ or $d' \le \epsilon = 2^{-l}$ for $l \ge n$, the KPA security problem does not arise under the qualification described in the last section. In addition, as discussed in section III and reinforced by [27], that is (exponentially) impossible to achieve with the usual key generation rate, and also cannot be achieved by privacy amplification. Furthermore, the key so generated in a concrete protocol is unlikely to be long enough to cover the message authentication key bits spent during key generation, such as in the case of [14]. See ref [2].

There is no ``unconditional security''
guarantee in QKD. Furthermore, we have now the following broader fundamental QKD security problems.

\begin{enumerate}[(i)]
\item There is no proof of security against known-plaintext attacks when
the generated key $K$ is used in direct encryption and Eve possesses
quantum memory, or in other composition security context.
\item The fundamental raw security level of Eve's probability of correctly estimating
any proper subset of $K$ is not bounded under either an $I_{ac}$ or $d$ constraint.
\item The true secure key rate is far smaller and is determined by
quantitative error exponents, the later rarely analyzed in security
proofs and for which $M$-ary quantum detection theory would be needed.
\end{enumerate}

With such difficulties on the foundation of quantum key distribution,
it appears radically new approaches are appropriate for fundamental security
guarantee.

\section*{ACKNOWLEDGEMENT}
I would like to thank Greg Kanter for his probing query, a main impetus
for this exposition, and for his comments. This work was supported
by the Air Force Office of Scientific Research.


\begin{thebibliography}{19}
\bibitem{key-15}In ref {[}2{]} the factor $\log(n)$ is missing in
quoting the $I_{E}$ value of ref {[}14{]}, as the more natural $I_{E}/n$
was taken to be the mutual information value instead of $I_{E}$ itself.
However, after Markov Inequality is taken into account in a natural
way {[}2,13{]} for individual instead of average guarantee, $l\sim10$
as quoted in ref {[}2{]}.
\end{thebibliography}
\end{document}